\documentstyle[12pt,epsfig]{article}

\begin{document}

\begin{center}
{\Large \bf Line emission from an accretion disk  around black hole: effects
of the disk structure}
\end{center}

\vspace{0.5cm}

\begin{center} 
{\large Vladimir I. Pariev$^{*\dagger}$, 
Benjamin C. Bromley$^{\ddagger}$}
\end{center}
\begin{center}
{\it $^*$Steward Observatory, University of Arizona, 933 N.Cherry Ave.,
Tucson, AZ 85721}\\
{\it $^\dagger$P.~N.~Lebedev Physical Institute, Leninsky Prospect 53,
Moscow 117924, Russia}\\
{\it $^{\ddagger}$MS-51, Harvard-Smithsonian 
Center for Astrophysics, 60~Garden Street, Cambridge, MA 02138}
\end{center}


\vspace{0.5cm}

{\bf Abstract.}
The observed iron $K\alpha$ fluorescence lines in Seyfert galaxies
provide strong evidence for an accretion disk near a supermassive
black hole as a source of the line emission.  Previous studies of line
emission have considered only geometrically thin disks, where the gas
moves along geodesics in the equatorial plane of a black hole.  Here
we extend this work to include effects on line profiles from finite
disk thickness, radial accretion flow and turbulence.  We adopt the
Novikov-Thorne solution, and find that within this framework,
turbulent broadening is the most significant effect.  The most
prominent changes in the skewed, double-horned line profiles is a
substantial reduction in the maximum flux at both red and blue peaks.
We show that at the present level of signal-to-noise in X--ray
spectra, proper treatment of the actual structure of the accretion
disk can change estimates of the inclination angle of the disk.  Thus
these effects will be important for future detailed modeling of high
quality observational data.

\section*{Introduction}

The {\it Advanced Satellite for Cosmology and Astrophysics} has
provided data from over a dozen Seyfert~\uppercase{i} galaxies to
reveal the presence of iron K$\alpha$ emission lines which are
broadened by a considerable fraction of the speed of light (greater
than 0.2~$c$ in some cases), and are consistent with relativistic thin
disk models\cite{mush95}, \cite{tanaka95}, \cite{nandra97} around a
Schwarzschild or Kerr black hole. One of the best studied galaxy is
MCG-6-30-15, which has been observed with highest signal to noise
ratio to show a highly broadened iron line from which disk model
parameters can be determined by data fitting.

During past few years more and more evolved models for physics of
K$\alpha$ line emission have been introduced as well as different
emissivity functions~\cite{dab97}, \cite{reybeg97}.  Despite this
progress the geometry and kinematics of the emitting material have
been left in an idealized state, unmodified since the work of
Cunningham 1975~\cite{cunn75}: the gas is limited to orbits in the
equatorial plane of the black hole; if the material is outside of the
radius of marginal stability, then the orbits are Keplerian, otherwise
the material free-falls onto the hole with the energy and angular
momentum of the innermost stable orbit.

The aim of present work is to consider effects of more realistic disk
geometry and kinematics on observed iron line profiles. For this
purpose we adopt the relativistic $\alpha$-disk model of Novikov \&
Thorne 1973~\cite{novthorne} and include explicitly the effects of
finite thickness, radial flow and turbulence. These processes can
significantly alter line profiles and must be modeled to make full use
of the high-quality data from upcoming space missions such as {\it AXAF},
{\it XMM}, and {\it ASTRO-E}.

\section*{Line profile calculation}

In order to evaluate additional Doppler and gravitational frequency 
shifts due to turbulent and radial velocities and disk thickness we adopt 
the standard thin disk model as given in \cite{novthorne} and
\cite{pagethorne}. The line is produced in the  inner part of the 
accretion disk, where radiation pressure dominates and opacity is
predominantly due to Thomson scattering.  Independent parameters which
determine the shape of the iron line are the specific dimensionless
angular momentum of a black hole $a=J/M$ (throughout we use units
where $c=1$ and $G=1$), the inclination angle of the disk $i$ ($i=0$
for a face-on geometry), the ratio of the disk luminosity to Eddington
luminosity $L/L_{edd}$, and the viscosity parameter $\alpha$.

Our assumptions in the model profile calculations are listed as
follows.  First, we assume that radiative efficiency of accretion is
equal to the binding energy per rest mass at the inner edge of the
accretion disk.  Within the disk itself, the source of emission is
located at the surface of a disk, where gas density
vanishes~\cite{novthorne}. In the frame which comoves with the bulk
flow, we assume that the emission is isotropic and monochromatic. For
random motions, we assume a Gaussian velocity spectrum, but consider
only the time-averaged effect on the line profile by convolution using
am r.m.s.~velocity given by the local sound speed in the disk.

To obtain a model line profile, we use a variant of
the geodesic solver described by Bromley, Chen
\& Miller 1997~\cite{brom97}. A pixel image of the observed disk
is generated to map frequency shifts over the surface of the disk.
Line profiles follow from binning pixels according to frequency,
with each pixel weighted by $g=\nu/\nu_e$ to the fourth power
(in accordance with the Liouville Theorem) and a factor given by
the local emissivity.

Line profiles are strongly dependent upon the assumed emissivity
law. In order to illustrate changes in line profiles introduced by
effects of the disk geometry and kinematics we choose a power law in
radius with index~2, i.e. $\epsilon(r)\propto r^{-2}$.  One can see
from Fig.~1 that the major effect of disk structure is broadening of
line profile and smoothing blue and red peaks.  The blue edge of the
line becomes softer and extends more toward the blue.  The overall
pointwise effect on line profiles at $L\simeq L_{edd}$ is about 10\%
or more, relative to the peak flux, which is about current level of
signal to noise for best measured iron line profiles.

\begin{figure} 
\vspace{-1.2in}
\centerline{\epsfig{file=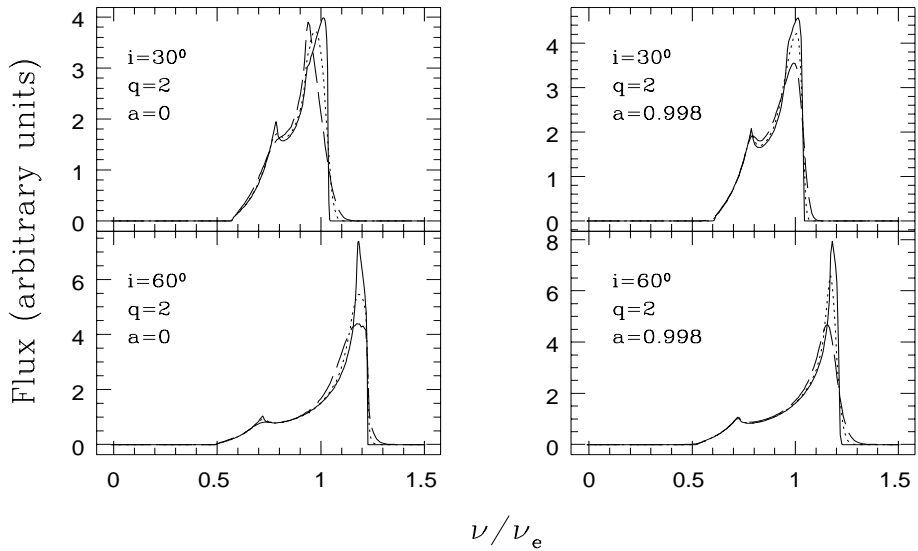,width=5.2in,height=4.8in}}
\vspace{10pt}
\vspace{-0.7in}
\caption{Line profile as a function of observed frequency $\nu$ (in units
of the emitted frequency $\nu_e$) for accretion disks in Schwarzschild
($a = 0$) and Kerr ($a = 0.998$) systems. Line profiles from
infinitesimally thin Keplerian disks are shown (solid curves) along
with profiles for turbulent disk with $L=L_{edd}/2$ and $\alpha=0.3$
(dotted lines), and for turbulent disk with $L=L_{edd}$, $\alpha=0.3$
(dashed lines).}\label{myfirstfigure}
\end{figure}
 
\section*{Application for interpretation of iron line observations}

In order to put constraints on inclination angle and radial location
of emitting material in a manner independent of the knowledge of
emissivity distribution over the disk surface we apply the method of
frequency extrema mapping~\cite{brompar} for the case of turbulent disk
with finite thickness.  In order to evaluate maximal effect of the
disk structure we calculated $g_{min}$--$g_{max}$ maps for a disk with
$L=L_{edd}$ and $\alpha=0.3$.  The main effect of turbulence is
increasing of $g_{max}$ and decreasing of $g_{min}$.

{\bf MCG-6-30-15.}  We consider three points in the
$g_{min}$--$g_{max}$ map from data given by Iwasawa et
al.~(1996)\cite{iwasawa96}, each corresponding to the disk in a
different phase of emissivity.  The ``intermediate'' and ``bright''
phases can be used in conjunction to constrain an inclination angle to
$i=22\pm 4$ degrees assuming $L=L_{edd}$, $\alpha=0.3$, and a Kerr
disk with $a = 0.998$ (lower than $i= 29\pm 5$ degrees for thin
Keplerian disk). With inclination angle constrains point from a ``deep
minimum'' phase allows to find limits on the location of the inner
emissive part of accretion disk to be within $R_{in}=5.3 \pm 0.8\, M$
(higher than $4.4\pm 0.8 \,M$ for thin Keplerian disk).  Similar
mapping for the case of a disk around a Schwarzschild black hole leads
to even lower estimate of inclination angle $i=13\pm 4$ degrees and
lower values for inner edge of emissive disk of $R_{in}=4.5\pm 0.7
\,M$.

{\bf Composite spectra.} We determine $g_{min}$ and $g_{max}$ for
spectra of 14 Seyfert~\uppercase{i} galaxies collected by Nandra et
al.~(1997)~\cite{nandra97}, for a single composite spectrum of the iron
$K\alpha$ line also presented in \cite{nandra97}, and for ``group A''
and ``group B'' of Seyfert~\uppercase{ii} galaxies from Turner et
al.~(1997)~\cite{turner97}.  In both extreme Kerr and Schwarzschild
cases with $L=L_{edd}$ and $\alpha=0.3$, we obtain lower upper
boundary for inclination angles of accretion disks $i<30^0$ in place
of $i<45^0$ for thin Keplerian disk. The inner radii of emission
change only a little and are about $8M$ for group~B objects and about
$20M$ for group~A objects. The effect of the black hole spin is in
decreasing of inner radius for group~B objects in Schwarzschild case
by about $2M$ with respect to Kerr case. Inclination angles and inner
radii are similar for Seyfert~\uppercase{i} and Seyfert~\uppercase{ii}
galaxies, thus, disfavoring unification scheme model.

\section*{Acknowledgments}

V.I.P. is pleased to thank S.~Colgate for discussions on many aspects
of this work and to the Theoretical Astrophysics Group of Los Alamos
National Laboratory for hospitality. B.C.B. acknowledges support from
NSF grant PHY~95-07695. The Cray supercomputer used in this research was
provided through funding from the NASA Offices of Space Sciences,
Aeronautics, and Mission to Planet Earth.

\end{document}